\documentclass[sigconf]{acmart}

\usepackage{amsmath}
\newcommand{\probP}{\text{I\kern-0.15em P}}
\AtBeginDocument{%
  \providecommand\BibTeX{{%
    \normalfont B\kern-0.5em{\scshape i\kern-0.25em b}\kern-0.8em\TeX}}}

\copyrightyear{2023} 
\acmYear{2023} 
\setcopyright{acmlicensed}\acmConference[WWW '23 Companion]{Companion Proceedings of the ACM Web Conference 2023}{April 30-May 4, 2023}{Austin, TX, USA}
\acmBooktitle{Companion Proceedings of the ACM Web Conference 2023 (WWW '23 Companion), April 30-May 4, 2023, Austin, TX, USA}
\acmPrice{15.00}
\acmDOI{10.1145/3543873.3584657}
\acmISBN{978-1-4503-9419-2/23/04}

\usepackage[inline]{enumitem}
\begin{document}

\title{CAM2: Conformity-Aware Multi-Task Ranking Model for Large-Scale Recommender Systems}

\author{Ameya Raul}
\authornote{Both authors contributed equally to this work.}
\affiliation{%
  \institution{Meta Inc.}
  \city{Menlo Park}
  \country{USA}
}
\email{araul@meta.com}

\author{Amey Porobo Dharwadker}
\authornotemark[1]
\affiliation{%
  \institution{Meta Inc.}
  \city{Menlo Park}
  \country{USA}
}
\email{ameydhar@meta.com}

\author{Brad Schumitsch}
\affiliation{%
  \institution{Meta Inc.}
  \city{Menlo Park}
  \country{USA}
}
\email{bschumitsch@meta.com}

\renewcommand{\shortauthors}{Raul, et al.}

\begin{abstract}
    Learning large-scale industrial recommender system models by fitting them to historical user interaction data makes them vulnerable to conformity bias. This may be due to a number of factors, including the fact that user interests may be difficult to determine and that many items are often interacted with based on ecosystem factors other than their relevance to the individual user. In this work, we introduce CAM2, a conformity-aware multi-task ranking model to serve relevant items to users on one of the largest industrial recommendation platforms. CAM2 addresses these challenges systematically by leveraging causal modeling to disentangle users' conformity to popular items from their true interests. This framework is generalizable and can be scaled to support multiple representations of conformity and user relevance in any large-scale recommender system. We provide deeper practical insights and demonstrate the effectiveness of the proposed model through improvements in offline evaluation metrics compared to our production multi-task ranking model. We also show through online experiments that the CAM2 model results in a significant 0.50\% increase in aggregated user engagement, coupled with a 0.21\% increase in daily active users on Facebook Watch, a popular video discovery and sharing platform serving billions of users.

\end{abstract}

\ccsdesc[500]{Information systems~Recommender systems}
\ccsdesc[500]{Computing methodologies~Ranking}
\ccsdesc[500]{Computing methodologies~Multi-task learning}

\keywords{Recommender system, Multi-task learning, Causal embedding, Conformity bias}

\maketitle

\section{Introduction}
Large-scale recommender systems provide personalized recommendations to users by considering their historical interactions and learning predictive models to fit that data. Such systems could result in strong conformity bias as they fail to take into account the fact that previous user interactions may have been influenced by multiple factors causing them to be inconsistent with user preferences. For example, a user may watch a video just because it has a lot of views, even if it doesn’t align with their true interests. This, in turn, can result in the system recommending more popular items rather than those that may be more relevant to the individual user, thereby limiting users’ exposure to long-tail content \cite{bobadilla2013recommender}.

Existing methods to disentangle conformity and interest address this primarily by eliminating popularity bias using a static global term on the item side \cite{bell2008bellkor}, while ignoring differing levels of conformity among users. To solve this issue, a recent work \cite{zheng2021disentangling} decomposes user interactions into two factors - conformity and interest, considering both users and items, and learns separate embedding representations for them with cause-specific data. This helps model the personalized conformance effect and captures the fact that different users have different levels of influence on their judgment. It computes user-item matching scores for both causes and sums them up to estimate the overall score on whether a user will interact with the item.

In this work, we present CAM2, a conformity-aware multi-task ranking model to improve user recommendations on Facebook Watch \footnote{https://www.facebook.com/watch} by extending the above idea of cause-specific embeddings. The main contributions of this work are as follows:
\begin{itemize}[leftmargin=*]
\item This work disentangles conformity and relevance based representations learning in the same model over the full training data by separating statistical interaction-based features from rich attribute-based and content-based features.
\item We present a novel loss formulation for our causal representation modules. In contrast to previous works that rely on aggregating all losses, we train these modules independent of the user interaction tasks' losses. We modify gradient back-propagation from the main network to each of the causal modules to ensure that the performance of the predicted user interaction tasks are optimized without ignoring the causal losses.
\item We provide deeper insights into efficiently using the personalized conformity-aware embeddings in our multi-task ranking model to achieve the best predictive performance across tasks.
\item Through experiments on our large-scale video recommendation platform, we validate the effectiveness of the proposed approach to improve prediction performance across user interaction tasks, leading to improvements in online metrics.
\end{itemize}

\section{Methodology}
\subsection{Problem Formulation}
Facebook Watch, with more than 1.25 billion monthly viewers \cite{FBWatchDap} is one of the largest global destinations for discovering and sharing video content. We use a typical two-stage recommendation ranking system design with a candidate generation stage and a ranking stage \cite{he2014practical, covington2016deep}. This paper focuses on the ranking stage, where the recommender has a few hundred promising candidates retrieved from the candidate generation stage. It applies a sophisticated, large-capacity model to rank these candidates and sort them in descending order of their relevance to users \cite{naumov2019deep}.

To effectively learn multiple types of user behavior and relevance dimensions, we use a multi-task deep neural network ranking model to predict multiple binary user interaction objectives such as user clicks, user interaction level with recommended video, user comments, etc. Similar to other multi-task ranking models, our system subsequently combines these predictions to compute a final relevance score for ranking.

Suppose we have a dataset of $n$ i.i.d. training data samples and $T$ is the number of user interaction tasks to model. All tasks share an input space $\{x_i\}$ consisting of user, item and user-item interaction features for example $i$ and $\{{y_i}^{t}\}$ is the binary user interaction label for the $t^{th}$ task for example $i$. The goal of our multi-task model is to learn a causal-aware scoring function $f(x,y,t|\theta)$ which predicts the probability of user interaction $t$ occurring on each example the model is applied on, parameterized by shared and task-specific parameters $\theta$. Our goal is to optimize predictions of the model for each interaction task objective by considering both conformity and relevance, to maximize recommendation performance metrics on future test data, that is not i.i.d. with the training data.

\subsection{Conformity-Aware Model Training}
\begin{figure}[h]
  \centering
  \captionsetup{font=small}
  \addtolength{\abovecaptionskip}{-7pt}
  \addtolength{\belowcaptionskip}{-10pt}
     \includegraphics[width=0.49\textwidth]{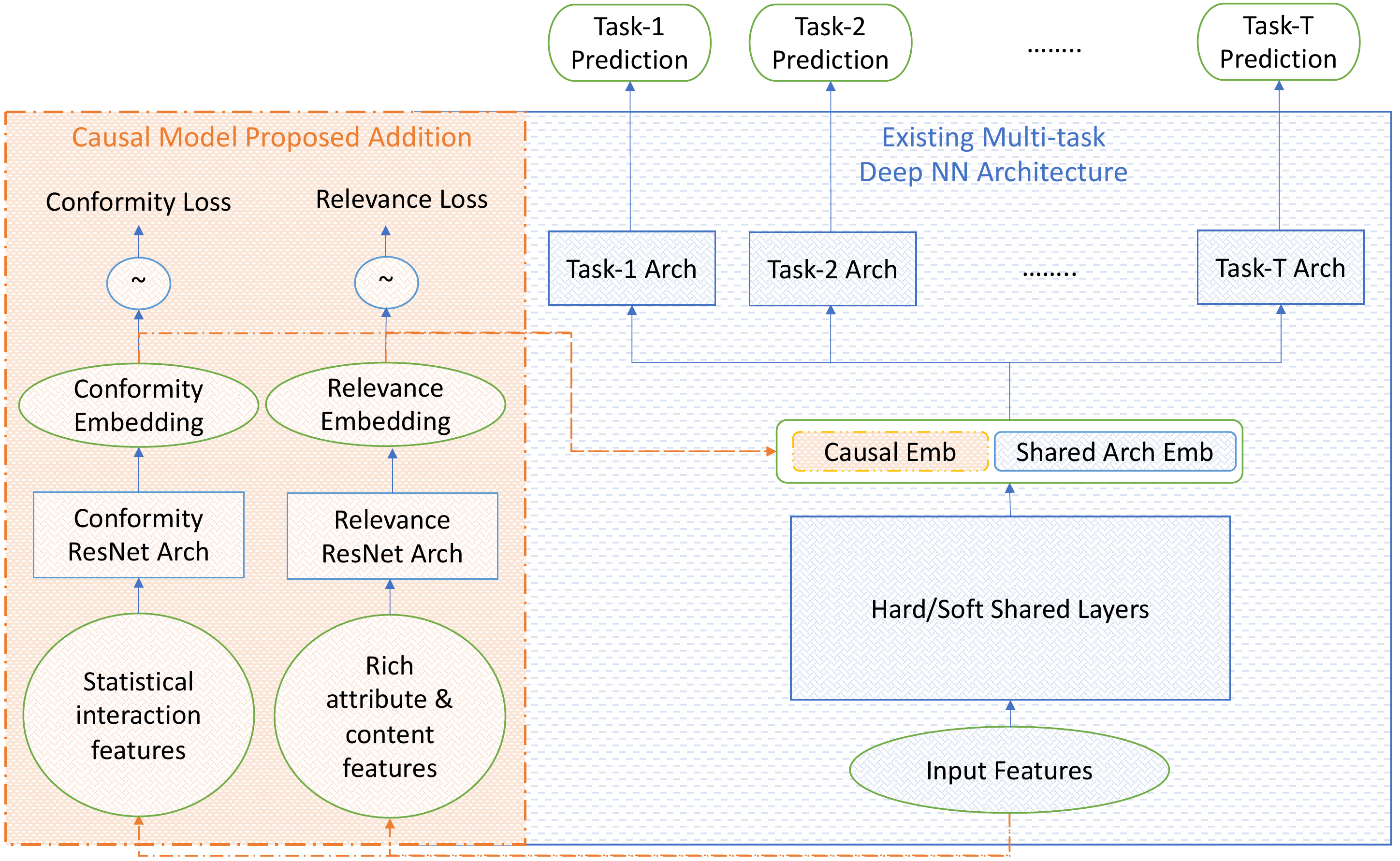}
     \caption{Proposed conformity-aware multi-task ranking model architecture with existing prod model architecture (blue box on the right) and proposed causal model addition to it (orange box on left).}
     \Description{Proposed conformity-aware multi-task ranking model architecture with existing prod model architecture shown with blue box on the right and proposed causal model addition to it in orange box on the left.}
     \label{fig:conformity-arch}
\end{figure}

Our existing multi-task deep neural network recommendation model serving production traffic on Facebook Watch consists of a sequence of multi-layer perceptrons, composed of a sequence of fully connected layers and an activation function applied component-wise. The model has a combination of hard and soft parameter sharing \cite{ruder2017overview} in the lower layers to exploit task relatedness better \cite{caruana1997multitask, baxter2000model} and enable some form of regularization as different tasks share model parameters. We input simple binary and continuous features directly into the model as real normalized values. In addition, embeddings are used to process the input of sparse categorical features into appropriate fixed-width dense representations. The model is trained with logistic regression under normalized cross-entropy (NE) loss. The blue box in Figure \ref{fig:conformity-arch} shows a simplified view of our existing production model architecture.

In order to improve relevance of recommendations by learning better representations, we need to better understand the underlying cause of user interactions. Our goal is to disentangle interactions due to user's alignment to conformity from those primarily due to user relevance that provide higher user value. We learn personalized causal embeddings in our multi-task ranking model corresponding to conformity and relevance based components using the steps outlined below. The orange box in Figure \ref{fig:conformity-arch} shows the proposed causal architecture addition to the production model.

\subsection{Causal Embeddings Model Architecture}
In this work, we train two separate cause-specific residual network (ResNet) modules \cite{he2016deep} as auxiliary tasks to learn decoupled causal embeddings indicating conformity alignment and relevance. We also introduce dedicated conformity and relevance losses to train these modules. They are only used to train the causal architectures for learning embedding representations jointly with the main model parameters, and not for model inference. We also disable gradient back-propagation from the task architectures to the causal modules to decouple the causal modules and prevent incorporating any bias in them from the task architectures. This allows us to train our causal embeddings on all available data without partitioning data into factor-specific data subsets based on the underlying reasons of user engagement.

First, the conformity module models the probability of user engagement due to conformity. There are two types of conformity:
\begin{itemize}[leftmargin=*]
\item User Conformity, denoted by $\overline{u}$, is the global likelihood that a user will engage with any given item.
\item Item Conformity, denoted by $\overline{i}$, is the global likelihood that an item will be engaged with by any user.
\end{itemize}

The conformity module predicts a combined conformity $\overline{c}$ by combining the losses from both of these types of conformity. The total conformity loss ($L_C$) is calculated as:
\[
  L_C(u, i) = \left\| (\overline{c} - \left\| \hat{u} + \hat{i} \right\|) \right\|
\]
where \begin{math}
    \left\| . \right\|
\end{math} denotes L2-norm and $\hat{u}$ and $\hat{i}$ are the predictions for user and item conformity respectively.

Second, the relevance module models the probability of a user engaging with an item based on the user's interest affinity with the item. Our model predicts whether the item is aligned with the user's interests or not. The total relevance loss ($L_R$) is calculated as:

\[
    L_R(u, i) = \sum_{x = 0}^{k} \left\| \overline{r_x} - (\hat{u_x} \cdot \hat{i_x}) \right\|
\]
where \begin{math}
    \left\| . \right\|
\end{math} denotes L2-norm, $\overline{r_x}$ denotes the true engagement rate of a user with interest $x$, while $\hat{u_x}$ is the predicted probability \textit{$\Pr(\text{User Engagement} \mid \text{Interest } x)$} and $\hat{i_x}$ is the predicted probability that the item belongs to interest $x$. Since each item can be associated with multiple interests $k$, we aggregate over all these interests to obtain the final loss for a user-item pair.

CAM2 predicts multiple user engagement tasks in the multi-task model, hence we take a weighted sum of these individual cause-specific losses along with our dedicated task losses to get the total loss $L$.
\[
  L = \sum_{t = 1}^{T} (w_t \cdot L_t) + w_C \cdot L_C + w_R \cdot L_R
\]
where $L_{t}$ is the loss for task $t$ and $w_{t}$ are tunable weights. To simplify the above formulation, we model this as a classification problem and define our causal labels by decomposing the probability of user engagement task $t$ into conformity and relevance as:
\begin{equation*}
\begin{split}
\Pr(t) & = w_1 \cdot \Pr(t \mid \text{Conformity}) + w_2 \cdot \Pr(t \mid \text{Relevance}) \\
& = w_1 \cdot \Pr(t \mid X >= thresh) + w_2 \cdot \Pr(t \mid X < thresh)
\end{split}
\end{equation*}
where $X$ is a scalar derived from user video historical engagement, used to distinguish the conformity and relevance separation with a static threshold $thresh$. Parameters $w_1$ and $w_2$ denote \textit{$\Pr(\text{Conformity})$} and \textit{$\Pr(\text{Relevance})$} respectively and are learnt jointly with the tasks. The conformity label is 1 if a user engagement happens and $X >= thresh$ and 0 otherwise, while the relevance label is 1 if user engagement happens and $X < thresh$ and 0 otherwise. Hence for the relevance loss, $\overline{r_x}$ is our label as defined above and our model predicts the product $(\hat{u_x} \cdot \hat{i_x})$. For conformity loss, $\overline{c}$ is our label as defined above and $\left\| \hat{u} + \hat{i} \right\|$ is our model's prediction.

This causal embedding training approach improves the main (non-auxiliary) tasks' predictions in the model, which are then combined to compute a final relevance score for ranking and recommending top-k items to the given user. The framework is generalizable and can be scaled to support multiple representations of conformity and user relevance using the above formulations.

\subsection{Input Features Partitioning}
The DICE framework \cite{zheng2021disentangling} disentangled representations for conformity and relevance by partitioning the training data into cause-specific parts, and trained different embeddings with cause-specific data. In contrast, we present a unique way to achieve this using all available training data samples by partitioning our features into two types:
\begin{itemize}[leftmargin=*]
\item Statistical engagement-based features such as number of video impressions, number of video views, social engagement rate of the user, click-through rate on the video, etc.
\item Rich attribute-based and content-based features such as user’s age, historical engaged videos, video interest topics, content type, video quality features, etc.
\end{itemize}

We hypothesize that statistical engagement-based features on both the video and user side are sufficient to learn user conformity alignment, while rich attribute-based and content-based features characterizing user and video properties enable user relevance learning. Combining these features results in a higher degree of entanglement and tends to skew the model towards popular and over-represented training data samples. Our input features partitioning approach allows the ranking model to learn more accurate embedding representations for both popular and long-tail groups by separating each cause of user engagement.

\section{Experiments}
\subsection{Experimental Setup}
We perform both offline and online evaluations on Facebook Watch, a real-world video recommendation system serving billions of users. We compare results against our production multi-task deep neural network ranking model, which is comparable to other large-scale industry video recommender ranking models \cite{covington2016deep}. The offline dataset contains 1.2B+ users, 100B+ instances and 35B+ user video engagements. It is split chronologically into the train set for model training and the next-day holdout test set for model performance evaluation. The models are trained recurrently on each day of additional data, starting with the previous network weights and optimizer state.

\subsection{Offline Experiments}
We use normalized cross-entropy (NE) as our primary offline evaluation metric as we expect accurate predictions, rather than just getting the optimal ranking order \cite{he2014practical}. The lower the NE value, the better the model prediction.

The basic conformity-aware multi-task ranking model as illustrated in Figure \ref{fig:conformity-arch}, consists of feature decomposition at the input layer, and two dedicated cause-specific conformity and relevance residual network (ResNet) modules trained with dedicated loss functions. The disentangled embeddings extracted from the corresponding ResNets are used to train the task architectures after concatenating with the main network’s shared bottom arch outputs. For simplicity, we refer to this variant as \textbf{CAM2-Proposed}. We conduct ablation studies to comparatively evaluate our modeling choices.
\begin{itemize}[leftmargin=*]
\item Embedding Usage: Instead of concatenating the embeddings with the shared bottom arch outputs, variant \textbf{CAM2-TaskArch} experiments with concatenating them directly at the final layer of the main engagement task architectures.
\item Causal Losses: In order to understand the importance of task independent causal losses, we experiment with a joint loss defined using a combination of existing engagement task labels and our causal labels in variant \textbf{CAM2-JointLoss}.
\item Input Features Partitioning: Variant \textbf{CAM2-AllFeats} experiments with adding all features to conformity and relevance modules. It serves as an ablation study on the input features partitioning we propose.
\end{itemize}

The ablation studies in Table~\ref{table:offline_results} show that our proposed CAM2 model that leverages causal embeddings to learn every task by concatenating with the shared bottom arch outputs, improves over the existing production model. Additionally, using input features partitioning and separate auxiliary conformity and relevance losses improve the model performance further. Our hypothesis is that leveraging disentangled representations in all stages of the model learning can encode different aspects of the reason for user engagement, resulting in better prediction performance across tasks.

\begin{table}
\captionsetup{font=small}
\addtolength{\abovecaptionskip}{-7pt}
\addtolength{\belowcaptionskip}{-6pt}
\begin{tabular}{c | c}
\toprule
Variant & Tasks NE (Aggregated)  \\
\midrule
CAM2-TaskArch       & -0.060\%              \\
CAM2-JointLoss       & -0.016\%             \\
CAM2-AllFeats       & +0.029\%              \\
\textbf{CAM2-Proposed}       & \textbf{-0.139}\%     \\
\bottomrule
\end{tabular}
\caption{Offline Results for ablation studies relative to production. Best results for the metric are indicated in bold.}
\label{table:offline_results}
\end{table}

\subsection{Online Experiments}
We deploy CAM2 on Meta's dedicated inference cloud \cite{hazelwood2018applied} and conduct three weeks of A/B testing to rank videos for users on Facebook Watch. The model predictions are used for scoring and generating a ranked list of videos displayed to users. The same ranking strategy and business logic is applied on control and treatment groups for fair comparison. Online A/B test shows that the causal model results in substantial wins across various metrics, all statistically significant with 95\% confidence interval.

\begin{figure}[h]
  \centering
  \captionsetup{font=small}
  \addtolength{\abovecaptionskip}{-7pt}
  \addtolength{\belowcaptionskip}{-7pt}
     \includegraphics[width=0.49\textwidth]{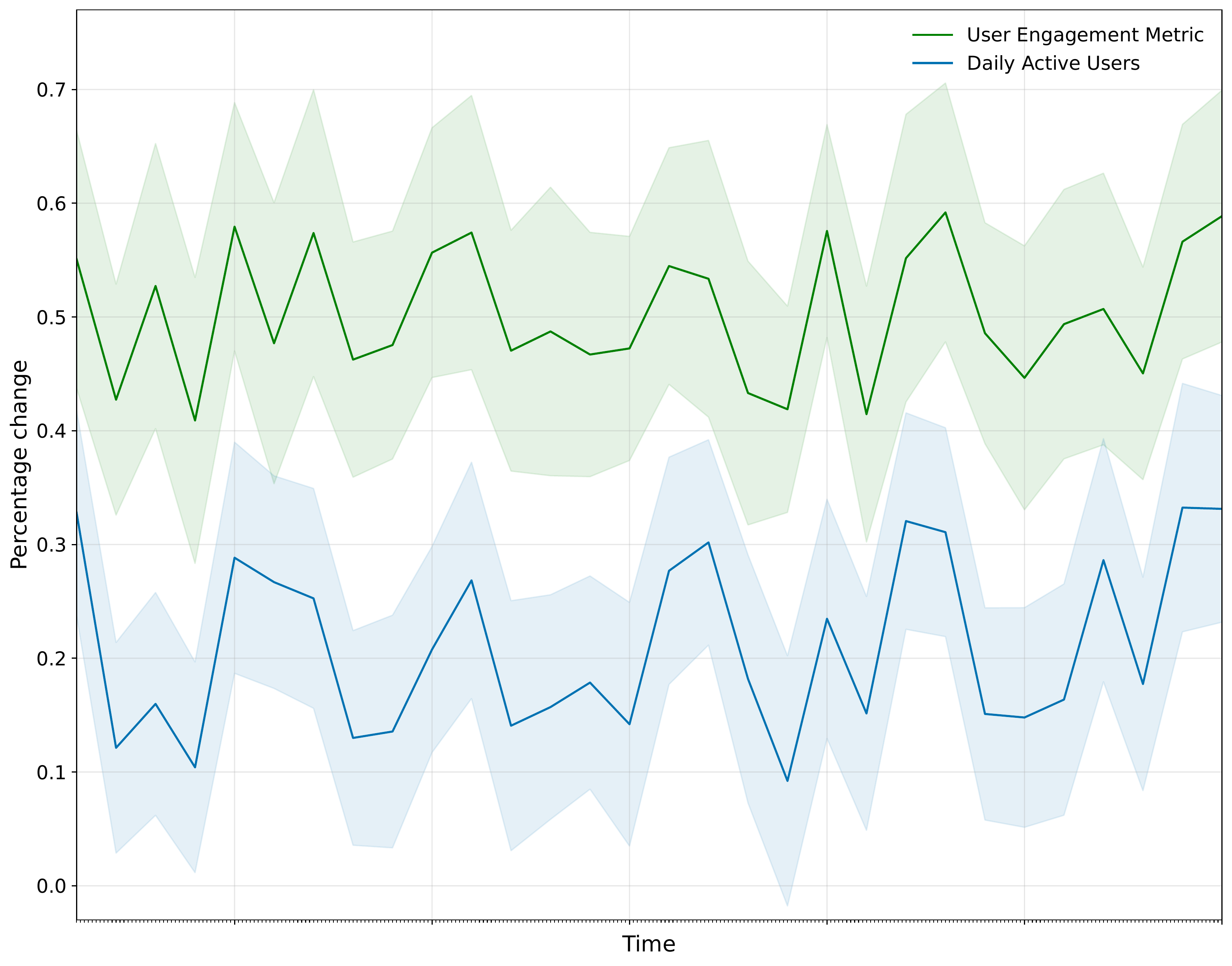}
     \caption{Overall user engagement metric and daily active users improvements on Facebook Watch in online A/B test.}
     \Description{Line chart showing overall user engagement metric and daily active users improvements on Facebook Watch in online A/B test, both statistically significant at 95\% confidence level.}
     \label{fig:online-engagement}
\end{figure}

Compared to production multi-task ranking model control, the treatment group using CAM2 for ranking videos in Watch shows $0.50\%$ increase in aggregated user engagement metric and $0.21\%$ increase in daily active users on Facebook Watch (Figure \ref{fig:online-engagement}), advancing our billion-scale video recommender system by a large margin.

\begin{table}
\captionsetup{font=small}
\addtolength{\abovecaptionskip}{-8pt}
\addtolength{\belowcaptionskip}{-6pt}
\begin{tabular}{c | c | c | c}
\toprule
\text{[0-1 day)} & \text{[1-3 days)} & \text{[3-10 days)} & \text{[10+ days)} \\ 
\midrule
$+2.43\%$  & $+1.17\%$  & $-0.28\%$  & Neutral              \\
\bottomrule
\end{tabular}
\caption{User engagement metric by video age relative to production.}
\label{table:online_age}
\end{table}

Our conformity-aware multi-task model reduces popularity bias by serving more long-tail and new videos. The number of videos resulting in $50\%$ and $75\%$ aggregated user engagements increased by $2.35\%$ and $2.06\%$ respectively. As shown in Table~\ref{table:online_age}, we also notice significant increase in user engagement metric on new videos compared to control. This demonstrates that the causal embedding can learn better long-tail representation and capture relevant user interests by separating conformity information, instead of recommending irrelevant popular videos.

Casual (low activity) users are defined as those users who have been active on Facebook Watch for 1 to 2 days in the last 28 days. To measure engagement as well as retention metrics on casual users, we freeze user activity levels just before the start of the experiment. Our model can better generalize to casual users by leveraging cause-specific embeddings, resulting in significant $0.83\%$ aggregated user engagement metric gains and $0.62\%$ increase in daily active users on Facebook Watch on this user cohort. This indicates that casual users have a better overall experience and are encouraged to return to the platform due to better recommendation quality.

\section{Related Work}
In recommender systems, it is important to consider that conformity can distort user ratings \cite{lederrey2018sheep}. When users are influenced by other factors in the ecosystem, their interactions may not be an accurate representation of their own interests \cite{liu2016you}. This can lead to biases in user ratings, and ultimately irrelevant recommendations. Existing approaches address this issue as an attempt to eliminate popularity bias through inverse propensity scoring (IPS) \cite{bottou2013counterfactual, gruson2019offline} or causal approaches \cite{wang2020causal, gupta2021causer, sinha2016deconvolving}. IPS-based methods re-weight popular items by the reciprocal of their popularity score to compensate for the fact that more popular items are more likely to be recommended. However, these methods are very sensitive to the weighting and don’t consider that not all popularity biases are bad, for example, viral and trending videos could be relevant to users and deserve to be recommended more. The causal methods mainly focus on confounding effects but lack fine-grained consideration of how to leverage popularity bias systematically to improve recommendation relevance.

Recent work on DICE \cite{zheng2021disentangling} which disentangles user conformity and interest representations is most relevant to our work. However, our proposed formulation of training personalized conformity-aware embeddings in the model as auxiliary tasks on the entire training data with input features partitioning, independent of the user interaction task losses has not been explored to the best of our knowledge.

\section{Conclusion}
In this work, we describe conformity bias as a real-world challenge in the design and development of large-scale recommender systems. We present a conformity-aware multi-task ranking model to address this issue and serve the most relevant recommendations. Our scalable and generalizable framework disentangles representations of user conformity and relevance in the model over the entire training data with input features partitioning and dedicated auxiliary losses. We demonstrate the effectiveness of our proposed method to improve user engagement and ecosystem value metrics on Facebook Watch, an industrial video discovery and sharing platform.

CAM2 develops a unique understanding of users by leveraging its input feature partitioning and loss formulation techniques. In the future, we will explore incorporating other user interaction causes in the model and investigate their impact on long-term metrics.

\bibliographystyle{ACM-Reference-Format}
\bibliography{sample-base}


\begin{thebibliography}{19}


\ifx \showCODEN    \undefined \def \showCODEN     #1{\unskip}     \fi
\ifx \showDOI      \undefined \def \showDOI       #1{#1}\fi
\ifx \showISBNx    \undefined \def \showISBNx     #1{\unskip}     \fi
\ifx \showISBNxiii \undefined \def \showISBNxiii  #1{\unskip}     \fi
\ifx \showISSN     \undefined \def \showISSN      #1{\unskip}     \fi
\ifx \showLCCN     \undefined \def \showLCCN      #1{\unskip}     \fi
\ifx \shownote     \undefined \def \shownote      #1{#1}          \fi
\ifx \showarticletitle \undefined \def \showarticletitle #1{#1}   \fi
\ifx \showURL      \undefined \def \showURL       {\relax}        \fi
\providecommand\bibfield[2]{#2}
\providecommand\bibinfo[2]{#2}
\providecommand\natexlab[1]{#1}
\providecommand\showeprint[2][]{arXiv:#2}

\bibitem[Baxter(2000)]%
        {baxter2000model}
\bibfield{author}{\bibinfo{person}{Jonathan Baxter}.}
  \bibinfo{year}{2000}\natexlab{}.
\newblock \showarticletitle{A model of inductive bias learning}.
\newblock \bibinfo{journal}{\emph{Journal of artificial intelligence research}}
   \bibinfo{volume}{12} (\bibinfo{year}{2000}), \bibinfo{pages}{149--198}.
\newblock


\bibitem[Bell et~al\mbox{.}(2008)]%
        {bell2008bellkor}
\bibfield{author}{\bibinfo{person}{Robert~M Bell}, \bibinfo{person}{Yehuda
  Koren}, {and} \bibinfo{person}{Chris Volinsky}.}
  \bibinfo{year}{2008}\natexlab{}.
\newblock \showarticletitle{The bellkor 2008 solution to the netflix prize}.
\newblock \bibinfo{journal}{\emph{Statistics Research Department at AT\&T
  Research}} \bibinfo{volume}{1}, \bibinfo{number}{1} (\bibinfo{year}{2008}).
\newblock


\bibitem[Bobadilla et~al\mbox{.}(2013)]%
        {bobadilla2013recommender}
\bibfield{author}{\bibinfo{person}{Jes{\'u}s Bobadilla},
  \bibinfo{person}{Fernando Ortega}, \bibinfo{person}{Antonio Hernando}, {and}
  \bibinfo{person}{Abraham Guti{\'e}rrez}.} \bibinfo{year}{2013}\natexlab{}.
\newblock \showarticletitle{Recommender systems survey}.
\newblock \bibinfo{journal}{\emph{Knowledge-based systems}}
  \bibinfo{volume}{46} (\bibinfo{year}{2013}), \bibinfo{pages}{109--132}.
\newblock


\bibitem[Bottou et~al\mbox{.}(2013)]%
        {bottou2013counterfactual}
\bibfield{author}{\bibinfo{person}{L{\'e}on Bottou}, \bibinfo{person}{Jonas
  Peters}, \bibinfo{person}{Joaquin Qui{\~n}onero-Candela},
  \bibinfo{person}{Denis~X Charles}, \bibinfo{person}{D~Max Chickering},
  \bibinfo{person}{Elon Portugaly}, \bibinfo{person}{Dipankar Ray},
  \bibinfo{person}{Patrice Simard}, {and} \bibinfo{person}{Ed Snelson}.}
  \bibinfo{year}{2013}\natexlab{}.
\newblock \showarticletitle{Counterfactual Reasoning and Learning Systems: The
  Example of Computational Advertising.}
\newblock \bibinfo{journal}{\emph{Journal of Machine Learning Research}}
  \bibinfo{volume}{14}, \bibinfo{number}{11} (\bibinfo{year}{2013}).
\newblock


\bibitem[Caruana(1997)]%
        {caruana1997multitask}
\bibfield{author}{\bibinfo{person}{Rich Caruana}.}
  \bibinfo{year}{1997}\natexlab{}.
\newblock \showarticletitle{Multitask learning}.
\newblock \bibinfo{journal}{\emph{Machine learning}} \bibinfo{volume}{28},
  \bibinfo{number}{1} (\bibinfo{year}{1997}), \bibinfo{pages}{41--75}.
\newblock


\bibitem[Covington et~al\mbox{.}(2016)]%
        {covington2016deep}
\bibfield{author}{\bibinfo{person}{Paul Covington}, \bibinfo{person}{Jay
  Adams}, {and} \bibinfo{person}{Emre Sargin}.}
  \bibinfo{year}{2016}\natexlab{}.
\newblock \showarticletitle{Deep neural networks for youtube recommendations}.
  In \bibinfo{booktitle}{\emph{Proceedings of the 10th ACM conference on
  recommender systems}}. \bibinfo{pages}{191--198}.
\newblock


\bibitem[Gruson et~al\mbox{.}(2019)]%
        {gruson2019offline}
\bibfield{author}{\bibinfo{person}{Alois Gruson}, \bibinfo{person}{Praveen
  Chandar}, \bibinfo{person}{Christophe Charbuillet}, \bibinfo{person}{James
  McInerney}, \bibinfo{person}{Samantha Hansen}, \bibinfo{person}{Damien
  Tardieu}, {and} \bibinfo{person}{Ben Carterette}.}
  \bibinfo{year}{2019}\natexlab{}.
\newblock \showarticletitle{Offline evaluation to make decisions about
  playlistrecommendation algorithms}. In \bibinfo{booktitle}{\emph{Proceedings
  of the Twelfth ACM International Conference on Web Search and Data Mining}}.
  \bibinfo{pages}{420--428}.
\newblock


\bibitem[Gupta et~al\mbox{.}(2021)]%
        {gupta2021causer}
\bibfield{author}{\bibinfo{person}{Priyanka Gupta}, \bibinfo{person}{Ankit
  Sharma}, \bibinfo{person}{Pankaj Malhotra}, \bibinfo{person}{Lovekesh Vig},
  {and} \bibinfo{person}{Gautam Shroff}.} \bibinfo{year}{2021}\natexlab{}.
\newblock \showarticletitle{CauSeR: Causal Session-based Recommendations for
  Handling Popularity Bias}. In \bibinfo{booktitle}{\emph{Proceedings of the
  30th ACM International Conference on Information \& Knowledge Management}}.
  \bibinfo{pages}{3048--3052}.
\newblock


\bibitem[Hazelwood et~al\mbox{.}(2018)]%
        {hazelwood2018applied}
\bibfield{author}{\bibinfo{person}{Kim Hazelwood}, \bibinfo{person}{Sarah
  Bird}, \bibinfo{person}{David Brooks}, \bibinfo{person}{Soumith Chintala},
  \bibinfo{person}{Utku Diril}, \bibinfo{person}{Dmytro Dzhulgakov},
  \bibinfo{person}{Mohamed Fawzy}, \bibinfo{person}{Bill Jia},
  \bibinfo{person}{Yangqing Jia}, \bibinfo{person}{Aditya Kalro},
  {et~al\mbox{.}}} \bibinfo{year}{2018}\natexlab{}.
\newblock \showarticletitle{Applied machine learning at facebook: A datacenter
  infrastructure perspective}. In \bibinfo{booktitle}{\emph{2018 IEEE
  International Symposium on High Performance Computer Architecture (HPCA)}}.
  IEEE, \bibinfo{pages}{620--629}.
\newblock


\bibitem[He et~al\mbox{.}(2016)]%
        {he2016deep}
\bibfield{author}{\bibinfo{person}{Kaiming He}, \bibinfo{person}{Xiangyu
  Zhang}, \bibinfo{person}{Shaoqing Ren}, {and} \bibinfo{person}{Jian Sun}.}
  \bibinfo{year}{2016}\natexlab{}.
\newblock \showarticletitle{Deep residual learning for image recognition}. In
  \bibinfo{booktitle}{\emph{Proceedings of the IEEE conference on computer
  vision and pattern recognition}}. \bibinfo{pages}{770--778}.
\newblock


\bibitem[He et~al\mbox{.}(2014)]%
        {he2014practical}
\bibfield{author}{\bibinfo{person}{Xinran He}, \bibinfo{person}{Junfeng Pan},
  \bibinfo{person}{Ou Jin}, \bibinfo{person}{Tianbing Xu}, \bibinfo{person}{Bo
  Liu}, \bibinfo{person}{Tao Xu}, \bibinfo{person}{Yanxin Shi},
  \bibinfo{person}{Antoine Atallah}, \bibinfo{person}{Ralf Herbrich},
  \bibinfo{person}{Stuart Bowers}, {et~al\mbox{.}}}
  \bibinfo{year}{2014}\natexlab{}.
\newblock \showarticletitle{Practical lessons from predicting clicks on ads at
  facebook}. In \bibinfo{booktitle}{\emph{Proceedings of the eighth
  international workshop on data mining for online advertising}}.
  \bibinfo{pages}{1--9}.
\newblock


\bibitem[Lederrey and West(2018)]%
        {lederrey2018sheep}
\bibfield{author}{\bibinfo{person}{Gael Lederrey} {and} \bibinfo{person}{Robert
  West}.} \bibinfo{year}{2018}\natexlab{}.
\newblock \showarticletitle{When sheep shop: measuring herding effects in
  product ratings with natural experiments}. In
  \bibinfo{booktitle}{\emph{Proceedings of the 2018 World Wide Web
  Conference}}. \bibinfo{pages}{793--802}.
\newblock


\bibitem[Liu et~al\mbox{.}(2016)]%
        {liu2016you}
\bibfield{author}{\bibinfo{person}{Yiming Liu}, \bibinfo{person}{Xuezhi Cao},
  {and} \bibinfo{person}{Yong Yu}.} \bibinfo{year}{2016}\natexlab{}.
\newblock \showarticletitle{Are you influenced by others when rating? Improve
  rating prediction by conformity modeling}. In
  \bibinfo{booktitle}{\emph{Proceedings of the 10th ACM conference on
  recommender systems}}. \bibinfo{pages}{269--272}.
\newblock


\bibitem[Naumov et~al\mbox{.}(2019)]%
        {naumov2019deep}
\bibfield{author}{\bibinfo{person}{Maxim Naumov}, \bibinfo{person}{Dheevatsa
  Mudigere}, \bibinfo{person}{Hao-Jun~Michael Shi}, \bibinfo{person}{Jianyu
  Huang}, \bibinfo{person}{Narayanan Sundaraman}, \bibinfo{person}{Jongsoo
  Park}, \bibinfo{person}{Xiaodong Wang}, \bibinfo{person}{Udit Gupta},
  \bibinfo{person}{Carole-Jean Wu}, \bibinfo{person}{Alisson~G Azzolini},
  {et~al\mbox{.}}} \bibinfo{year}{2019}\natexlab{}.
\newblock \showarticletitle{Deep learning recommendation model for
  personalization and recommendation systems}.
\newblock \bibinfo{journal}{\emph{arXiv preprint arXiv:1906.00091}}
  (\bibinfo{year}{2019}).
\newblock


\bibitem[Rajwat(2020)]%
        {FBWatchDap}
\bibfield{author}{\bibinfo{person}{Paresh Rajwat}.}
  \bibinfo{year}{2020}\natexlab{}.
\newblock \bibinfo{title}{The Evolution of Facebook Watch}.
\newblock
\newblock
\urldef\tempurl%
\url{https://about.fb.com/news/2020/09/the-evolution-of-facebook-watch/}
\showURL{%
Retrieved Oct 27, 2022 from \tempurl}


\bibitem[Ruder(2017)]%
        {ruder2017overview}
\bibfield{author}{\bibinfo{person}{Sebastian Ruder}.}
  \bibinfo{year}{2017}\natexlab{}.
\newblock \showarticletitle{An overview of multi-task learning in deep neural
  networks}.
\newblock \bibinfo{journal}{\emph{arXiv preprint arXiv:1706.05098}}
  (\bibinfo{year}{2017}).
\newblock


\bibitem[Sinha et~al\mbox{.}(2016)]%
        {sinha2016deconvolving}
\bibfield{author}{\bibinfo{person}{Ayan Sinha}, \bibinfo{person}{David~F
  Gleich}, {and} \bibinfo{person}{Karthik Ramani}.}
  \bibinfo{year}{2016}\natexlab{}.
\newblock \showarticletitle{Deconvolving feedback loops in recommender
  systems}.
\newblock \bibinfo{journal}{\emph{Advances in neural information processing
  systems}}  \bibinfo{volume}{29} (\bibinfo{year}{2016}).
\newblock


\bibitem[Wang et~al\mbox{.}(2020)]%
        {wang2020causal}
\bibfield{author}{\bibinfo{person}{Yixin Wang}, \bibinfo{person}{Dawen Liang},
  \bibinfo{person}{Laurent Charlin}, {and} \bibinfo{person}{David~M Blei}.}
  \bibinfo{year}{2020}\natexlab{}.
\newblock \showarticletitle{Causal inference for recommender systems}. In
  \bibinfo{booktitle}{\emph{Fourteenth ACM Conference on Recommender Systems}}.
  \bibinfo{pages}{426--431}.
\newblock


\bibitem[Zheng et~al\mbox{.}(2021)]%
        {zheng2021disentangling}
\bibfield{author}{\bibinfo{person}{Yu Zheng}, \bibinfo{person}{Chen Gao},
  \bibinfo{person}{Xiang Li}, \bibinfo{person}{Xiangnan He},
  \bibinfo{person}{Yong Li}, {and} \bibinfo{person}{Depeng Jin}.}
  \bibinfo{year}{2021}\natexlab{}.
\newblock \showarticletitle{Disentangling user interest and conformity for
  recommendation with causal embedding}. In
  \bibinfo{booktitle}{\emph{Proceedings of the Web Conference 2021}}.
  \bibinfo{pages}{2980--2991}.
\newblock


\end{thebibliography}

\end{document}